\documentclass[a4paper,12pt]{article}

\usepackage[english]{babel}
\usepackage[dvips]{graphicx}
\usepackage{amsmath}
\usepackage{amssymb}

\begin{document}

\title{Exact Hurst exponent and crossover behavior in a limit order market 
model}
\author{R.D. Willmann$^{1,2}$, G.M. Sch\"utz$^2$ and D. Challet$^3$\\
\footnotesize
$^1$ Department of Physics of Complex Systems, \\
\footnotesize
Weizmann Institute of Science, 76100 Rehovot, Israel \\
\footnotesize
e-mail: fewillma@wicc.weizmann.ac.il \\
\footnotesize
$^2$ Institut f\"ur Festk\"orperforschung, Forschungszentrum J\"ulich, 52425 J\"ulich, Germany \\
\footnotesize
$^3$ Theoretical Physics, 1 Keble Road, Oxford OX1 3NP, United
Kingdom}
\maketitle
\begin{abstract}
An exclusion particle model is considered as a highly simplified model of a
limit order market. Its price behavior reproduces the well known crossover 
from over-diffusion (Hurst exponent $H>1/2$) to diffusion ($H=1/2$) when the time horizon is 
increased, provided that orders are allowed to be canceled. For early times 
a mapping to the totally asymmetric exclusion process yields the exact result 
$H=2/3$ which is in good agreement with empirical data. The underlying
universality class of the exclusion process suggests some robustness of the 
exponent with respect to changes in the trading rules. In the crossover 
regime the Hurst plot has a scaling property where the bulk deposition/cancellation rate is 
the critical parameter. Analytical results are fully supported by numerical 
simulations.
\end{abstract}


\section{Introduction}

Financial markets have in recent years been at the focus of physicists 
attempts to apply existing knowledge from statistical mechanics to economic 
problems \cite{Stanley,Bouchaud,FarmerIvory,Daco}. These markets,
though largely varying in 
details of trading rules and traded goods, are characterized by 
some generic features of their financial time 
series, called `stylized facts' \cite{Stanley,Bouchaud,Daco}. 
Agent based models of financial markets are successful to reproduce some 
stylized facts \cite{Cont,Lux,J99,CCMZ00,CMZ01,BouchMG}, such as volatility 
clustering,  fat-tailed probability distribution of price increments and 
over-diffusive price behavior at short time scales and diffusive behavior 
at later times. But all of them need an explicit price formation rule that 
links excess demand to price changes \cite{FarmerPrice,Cont,J00,Helbing}, 
that can be itself problematic. An other approach consists in modeling the 
price formation, for instance in limit order markets 
\cite{Bak,Maslov,CS01,Farmer,BouchLimit}. So far, all these  models of limit 
order markets have under-diffusive prices at short times, with a
crossover to diffusive prices at longer times for some of
them. Underdiffusive behavior at short times is realistic in limit
order markets, but all these models lack the over-diffusive price behavior observed in real
markets at medium time scales.
Here, we introduce a crude non-equilibrium model with overdiffusive
price that is able to reproduce the 
crossover from a Hurst exponent $H=2/3$ to $H=1/2$ at larger times, when
correlations in the price dynamics are washed out by cancellations of existing
orders and independent placements of new orders. In the early time regime
our model belongs to the $1d$-KPZ universality class \cite{kpzuniversal}, hence, 
its mechanism for over-diffusive price spreading is robust and analytically 
tractable.

In section II we define our model in terms of the limit order market dynamics. 
Our simulation results are presented in section III. In section IV the
equivalence of the early-time regime of our model to the totally asymmetric 
exclusion process (TASEP) \cite{tasep} with a second class particle is
established and its relation to the KPZ \cite{kpz} and noisy Burgers 
\cite{Spohn} equation are discussed.

\section{Model definition}

We consider two types of orders: {\em limit orders} that are wishes to buy or 
sell a given quantity of stock at a given price, and {\em market orders}
that are orders to immediately sell or buy an asset at the best 
instantaneous price. Limit orders are stored in an order book until they are 
either canceled,\footnote{For instance because they had a predefined maximal 
lifetime.} or fulfilled, provided that the current market price has moved 
towards their  prices. 
The model is constructed as a one dimensional lattice model, in a similar 
spirit as in \cite{Bak,Tang}. Let the lattice of length $L$ represent the 
price axis, with the lowest price on the left end at site 1 and the highest 
one on the right at site $L$. Limit orders of the two different kinds, i.e., 
asks ($A$) and bids ($B$) are placed on the lattice according to the price 
that the order is based on. As bids name lower prices than asks, they will be 
found on the left side of the lattice. The current market price ($x$) 
separates the two regions. Sites representing prices at which currently no 
order is placed are indicated as $0$. 

In the model we made the following 
simplifying assumptions: 
\begin{itemize}
\item Only one kind of asset is traded and its price dynamics is not 
directly influenced by outside sources but just by the state of its limit 
order book.
\item Each site can only carry one order (exclusion model).
\item Limit orders of either kind come in a unit size.
\item Only a finite price interval of width $L$ is considered.
\end{itemize}
The last assumption is justified as trade only takes place in a narrow 
interval around the market price. In our model $L$ can be chosen arbitrarily.
For a discussion on differences between models and real limit order markets 
see e.g. \cite{Mills}.  

\begin{figure}[h]
\begin{center}
\includegraphics[scale=1.0]{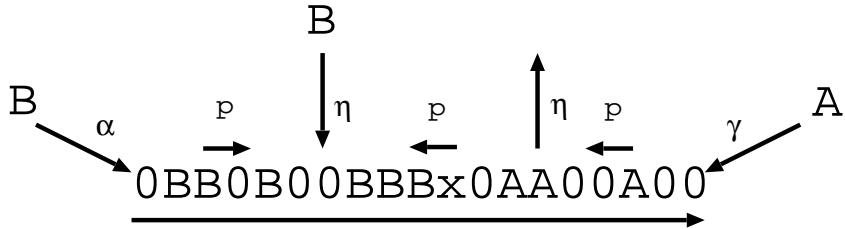}
\caption{Example of a configuration and possible moves with their assigned rates.}
\end{center}
\end{figure}

The dynamics of the lattice is as follows (see Figure 1):
\begin{itemize}
\item At site $1$ bids enter the system at rate $\alpha$: $0 \rightarrow B$.
\item At site $L$ asks enter the system at rate $\gamma$. $0 \rightarrow A$.
\item Asks and bids can diffuse one site towards the market price at rate $p$ 
provided no other order  is already placed at the target site: $B0 \rightarrow
 0B, \enspace 0A \rightarrow A0$.
\item Bids can be placed at unoccupied sites left of the market price at rate 
$\eta$. The same holds for asks being placed right of the market price: 
$0 \rightarrow A, \enspace 0 \rightarrow B$.
\item Bids and asks can be evaporated at rate $\eta$: $A \rightarrow 0, 
\enspace B \rightarrow 0$. This reflects both orders being canceled as well 
as being timed out. We have made the simplifying assumption of a constant 
removal rate at each site instead of considering the lifetimes of individual 
orders.
\item An ask can be fulfilled at rate $p$ by an incoming market order, 
provided it is adjacent to the current market price. Thus the order is 
removed: $xA \rightarrow 0x$.
\item A bid can be fulfilled at the same condition and rate, leading to a 
decrease of the market price and removal of the order: $Bx \rightarrow x0$.
\end{itemize}
The role of order injection and diffusion is to ensure a fluctuating order 
density on both sides of the price. On the other hand, the dynamics of the 
special particle which represents the price, is such that the sign of price 
increments is constant as long as the bid-ask spread is not minimal, i.e., as 
long as the price is not surrounded by two orders. This is a crude but 
efficient way of implementing trends in limit order market models. Notice that
this dynamics implies that the price is always between the best bid and best 
ask orders, which is true 95\% of the time in ISLAND ECN (www.island.com). 
Even if it is likely that orders do not diffuse \cite{CS01}, we use this 
ingredient as a way of obtaining exact results for the Hurst 
exponent.

\section{Simulation results}

Throughout our simulations we chose initial configurations where each site of 
the lattice is randomly occupied by an order with probability $1/2$. 
Furthermore we chose $\alpha=\gamma=1/4$, $p=1/2$. The lattice size $L$ was 
chosen such that the market price $M$ could not fluctuate out of the 
represented price interval during the simulation time. The choice of rates 
guarantees that the price has no drift but on average remains on its initial 
position, i.e., $L/2$. Averaging over initial conditions is implied in all 
our simulations.

In the case $\eta=0$ we remain with a model where price dynamics is solely 
caused by diffusion of limit orders. We are mainly interested in the Hurst 
exponent $H$ defined by the relationship 
$\left<(\delta x)^2\right>^{1/2}\propto\Delta t^H$. 

\begin{figure}[h]
\begin{center}
\includegraphics[scale=0.5]{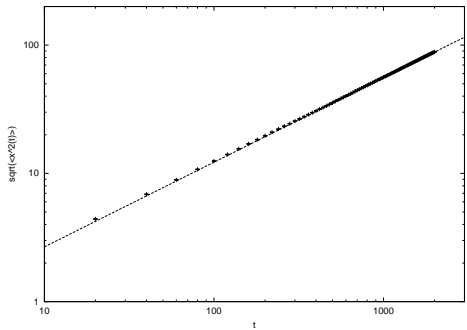}
\caption{Fluctuations of the price signal 
$\sqrt{\langle x^2(t) \rangle} \propto t^H$ at $\eta=0$ and fit with $H=2/3$.}
\end{center}
\end{figure}

In Fig. 2 we show
the fluctuation of the price position $x$ relative to the initial price
versus time, in a double-logarithmic plot. As can be seen the Hurst exponent 
of the model is $H=2/3$ for all times without a crossover. This 
behavior is in contrast to the corresponding result from the basic Bak, 
Paczuski, Shubik (BPS) model \cite{Bak,Tang}. In the BPS models offers to 
buy and sell diffuse on a lattice representing prices just as in our model. 
The difference is that upon meeting offers to sell and buy mutually annihilate
in the BPS model, thus carrying the model to the universality class of the 
$A+B \rightarrow 0$ reaction diffusion model. For that model it is known 
analytically that $H=1/4$ at long times plus logarithmic corrections at 
shorter times \cite{Barkema}. In our model no mutual annihilation 
(of ask and bid), takes place,
but just one type of order vanishes (ask or bid, fulfilled together with a
market order), thus causing a price change. This carries our 
model into the realm of the KPZ universality class as we will illustrate in 
the next section and yields $H=2/3$. 

\begin{figure}[h]
\begin{center}
\includegraphics[scale=0.5]{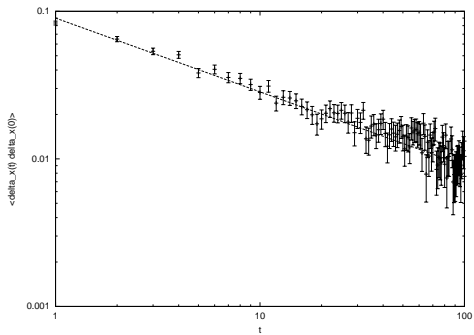}
\caption{Correlations of the increments at $\eta=0$ and fit by the eye with a 
function decaying algebraically as $t^{-1/2}$.}
\end{center}
\end{figure}

Price increments $\delta x(t)=x(t'+t)-x(t')$ show algebraically decaying 
correlations (Figure 3), with $\langle \delta x(t') \enspace \delta x(t'+t) 
\rangle \propto t^{-1/2}$ whereas these correlations should be essentially 
zero; this is due to the absence of evaporation (see below); the correlation 
of absolute increments has algebraically decreasing autocorrelation 
with an exponent of approximately 1. These long ranged correlations cause 
the over-diffusive behavior. 

\begin{figure}[h]
\label{bild10}
\begin{center}
\includegraphics[scale=0.5]{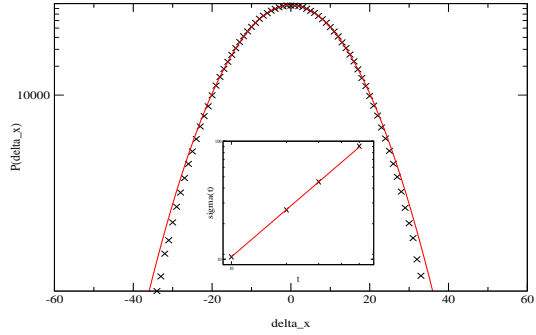}
\caption{Distribution of the increments (logarithmic scale) for $t=50$ at 
$\eta=0$ and Gaussian fit. The inset shows the variance of the distribution 
as a function of time and a fit proportional to $t^{4/3}$.}
\end{center}
\end{figure}

Due to the price process itself 
($|\delta x(1)|\in{0,1}$), the histogram of 
$\delta x(t)$ is almost Gaussian in shape, the tails appear even less 
pronounced than a Gaussian pdf (Figure 4). The variance of the distribution 
of increments $\delta x(t)$ increases as $\sigma \propto t^{4/3}$ (see inset 
of Figure 4). The dynamical
exponent of the price process extracted from this property is $z=3/2$.
Clearly the stochastic process causing the price 
movements is not Gaussian, not even a rescaled Gaussian. 

\begin{figure}[h]
\begin{center}
\includegraphics[scale=0.5]{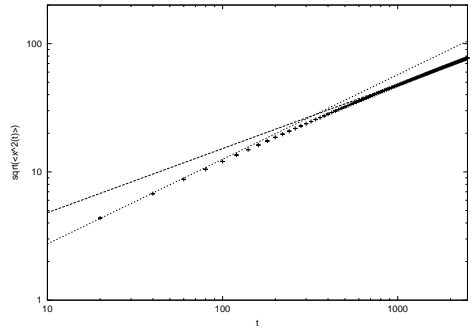}
\caption{Fluctuations of the price signal $\sqrt{\langle x^2(t) \rangle} 
\propto t^H$ at $\eta=1/512$ and fit with $H=2/3$ in the early time regime 
and $H=1/2$ at late times.}
\end{center}
\end{figure}

In what follows we consider the case $\eta \neq 0$. Clearly this is more 
realistic than $\eta = 0$ as it is possible to place orders of either 
kind at any 
unoccupied site on the price axis without having to perform diffusion steps 
all the way from the boundaries. Also the withdrawal of orders due to 
cancellation and timeout becomes thus possible. As seen in Figure 5 the 
fluctuations of the price signal show a crossover from
super-diffusive behavior at short times, characterized by the Hurst 
exponent $H=2/3$ to diffusive behavior at later times, implying $H=1/2$. 

\begin{figure}[h]
\begin{center}
\includegraphics[scale=0.5]{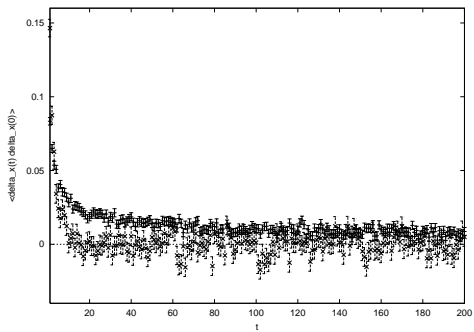}
\caption{Decay of the correlations of the increments for $\eta=0$ (upper curve)
and $\eta=1/8$ (lower curve).}
\end{center}
\end{figure}

The local random events controlled by the parameter $\eta$ destroy the long 
ranged time correlations of the price increments. This can be seen in 
Figure 6, showing the correlation of increments $\langle \delta x(t) 
\enspace \delta x(0) \rangle$ as a function of $t$ for $\eta=0$ and 
$\eta=1/8$. At $t=200$ the correlation function for the $\eta=0$ case is 
still different from zero, while a decay to zero for the other case occurred 
long since. Note that the autocorrelation of price increments should be 
negative for short time whereas it is positive in our model; this is due to 
the fact that we do not allow the coexistence of the two types of markets 
orders (see \cite{CS02}).

\begin{figure}[h]
\begin{center}
\includegraphics[scale=0.5]{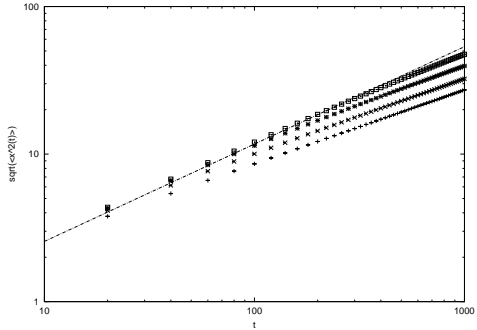}
\caption{Fluctuations $\sqrt{\langle x^2(t) \rangle }$ for $\eta=1/8$, 
$\eta=1/32$, $\eta=1/128$ and $\eta=1/512$ (from below). The parameter 
$\eta$ controls the range over which $H=2/3$ (dashed line).}
\end{center}
\end{figure}

Adjusting the rate $\eta$ serves as a parameter to control the crossover 
point. Figure 7 shows the price fluctuations for values of $\eta$ between 
1/8 and 1/512. The larger the rate $\eta$ for local events, the shorter is 
the time span for over-diffusive fluctuations. In fact, in our simulations 
over-diffusive behavior over a longer time 
interval appears only to be possible at seemingly meaningless low rates
$\eta$, namely $\eta \approx 1/L$, compared to $\alpha=\gamma=1/4$ and $p=1/2$.
From the study of empirical data of the Island ECN limit order market 
conducted in \cite{CS01} it is known that about 80 per cent of limit orders 
in the respective market vanished due to timeouts. Only 20 per cent of the 
offers were (at least partially) fulfilled. We measured these quantities as a function 
of $\eta$ in our simulations, where fulfillment of an order means either the 
process $Bx \rightarrow x0$ or $xA \rightarrow 0x$ and timeouts are reflected 
by the rate $\eta$. It turns out that for a lattice of $L=1000$ at 
$\eta=1/512$ about 8 per cent of orders were fulfilled and at $\eta=1/1024$ 
about 16 per cent. Thus the choice of small $\eta$ matching the observed 
fulfillment rate is realistic and could in fact be used as a means to gauge 
the simulation time by comparing the known empirical crossover time and the 
simulation crossover.

In the spirit of dynamical scaling it is tempting to assume that the price 
fluctuations with sufficiently low $\eta$ can be described in terms of a 
scaling function $F$ with
\begin{equation*}
<x^2(t)>=\eta^{-\mu_1}F(\eta^{\mu_2} t)
\end{equation*}
Since $\eta$ is a rate with inverse dimension of time one expects
$\mu_2=1$ for covariance of $<x^2(t)>$ under rescaling of time.
For small times, i.e., small arguments of the scaling function, this ansatz
should reproduce the behavior $<x^2(t)> \propto t^{4/3}$ which implies
$F(y) \propto y^{4/3}$ for $y  \to 0$. Independence of $\eta$ thus yields
the scaling relation
\begin{equation}
\label{scalingrel}
4/3 \mu_2 = \mu_1
\end{equation}
Hence one expects $\mu_1=4/3$.
For large times crossover to diffusion implies $F(y) \propto y$ for 
$y  \to \infty$. For large $\eta$ (of order 1 and larger) we obtain ordinary 
random walk behavior even at early times and the scaling relations are not
expected to be valid.

\begin{figure}[h]
\begin{center}
\includegraphics[scale=0.5]{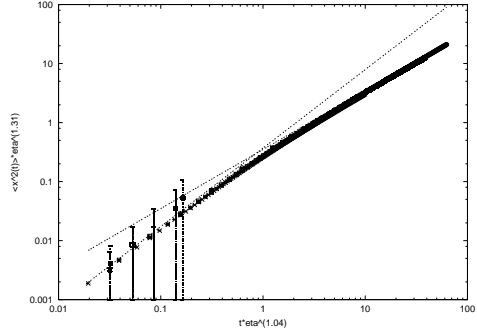}
\caption{Data collapse for the data for $\eta=1/32$, to $\eta=1/1024$ using 
the scaling function given in the text and lines to guide the eye 
corresponding to $H=2/3$ and $H=1/2$.}
\end{center}
\end{figure}

These arguments are well born out by Monte Carlo simulations. The best fit 
for the data could be achieved for the choice $\mu_1=1.31$ and $\mu_2=1.04$. 
These exponents are used in the plot (Fig. 8). 
The scaling property suggests that the relevant time scale of the model is 
$\tau=1/ \eta$, which is the average time between successive placements or 
evaporations of an order at a given site.

\begin{figure}[h]
\begin{center}
\includegraphics[scale=0.5]{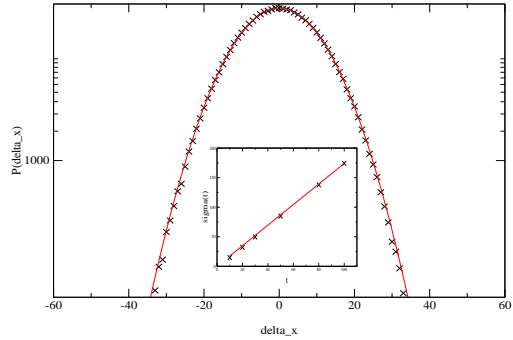}
\caption{Distribution of the increments for t=50 at $\eta=1/8$ and Gaussian 
fit. The inset shows the variance of the distribution as a function of time 
and the linear asymptote.}
\end{center}
\end{figure}

The increments of the price signal $\delta x(t)$ for $\eta \neq 0$ (Fig. 9) 
differ from $\eta=0$ in two important respects. Firstly, the tails
of the distribution are closer to a Gaussian. A second and more pronounced
difference is the behavior of the variance $\sigma$ of 
the distribution as a function of time, which shows a crossover from 
$\sigma \propto t^{4/3}$ to $\sigma \propto t$ just as the price signal 
itself. This means that the price performs an ordinary random walk at long 
times. A snapshot of the price movement is shown in Figure 10.

\begin{figure}[h]
\begin{center}
\includegraphics[scale=0.5]{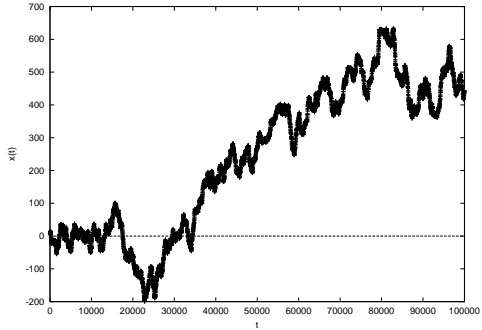}
\caption{Snapshot of price movement at $\eta=1/8$.}
\end{center}
\end{figure}

\section{Connection to the TASEP}

The virtue of our model is the equivalence at $\eta=0$ to the totally 
asymmetric exclusion process (TASEP) for which a wealth of exact results 
exists \cite{tasep}. In the TASEP excluding particles enter a lattice at rate 
$\alpha$ from the left and hop with rate $p$ to the right, provided the 
target site is empty. At the right end they can leave the system with rate 
$\beta$. In connection with the TASEP a 'second class' particle 
\cite{Ferrari} has been defined to have the following dynamics: A first class
particle meeting upon a second class particle to its right will exchange 
places. A second class particle with a vacant right neighboring site hops to 
that site. The second class particles motion is designed such that it does not
interfere with the motion of the first class particles. In fact, the motion 
of a single second class particle in the system is on average that of a 
density fluctuation in the system.

Upon coarse graining the dynamics of the TASEP can be described by the noisy 
Burgers equation, which is closely related to the KPZ equation known to have 
a universal dynamical exponent $z=3/2$ \cite{kpzuniversal}. This implies a
Hurst exponent $H=2/3$ as discussed above. For the noisy Burgers equation 
the over-diffusive spreading of a density fluctuation (i.e., the spreading of 
the second class particle, representing the price signal in our model) with 
$H=2/3$ has been shown analytically \cite{Spohn} in the case of statistically 
averaging over initial positions as well as realizations, which is always 
implied in our simulations. 

The mapping between our model and the TASEP is as follows: The market price 
$x$ represents the second class particle. Left of its position, bids are first
class particles in the sense of the TASEP and vacancies remain what they are. 
To the right of the market price vacancies take the role of first class 
particles in the TASEP sense and asks that of vacancies. The price dynamics in
our model is precisely that of a second class particle or density fluctuation 
in the TASEP. 

The TASEP may be 
seen as a discretized version of the noisy Burgers equation. It is an
exactly solvable model for which $z=3/2$ has been obtained through
the Bethe ansatz \cite{Gwa92}. More recently, also the distribution
of the second class particle for averaged random initial conditions has been
calculated exactly through a correspondence with statistical properties
of random matrices \cite{Prae01}. This confirms the results of our
simulations for a finite lattice with open boundaries, but system
large enough to be equivalent to an infinite system. We have also
performed simulations for a {\em fixed} random initial state. We found
that the super-diffusive spreading of the second class particle prevails,
but the amplitude $<x^2>/t^{4/3}$ depends on the initial condition.
This is in accordance with expectations \cite{Spohn01}.

\section{Conclusions}

In this work we have presented a model exhibiting the empirically observed 
crossover of the Hurst exponent from $H>1/2$ to $H=1/2$. By a mapping to
the totally asymmetric exclusion process we obtain the exact value $H=2/3$ 
which is close to what is often observed in real markets. 
The existence of an exact analytic solution puts our model in contrast to the 
model by Bak \cite{Bak} exhibiting over-diffusive spreading by volatility 
feedback into the system and a copying strategy of the traders, but for which 
no analytical solution is known.
The over-diffusive behavior results from time correlations build up in the
biased internal motion of asks and bids respectively which together with
market orders drive the price process. We identify the average time 
between evaporation events of orders (due to time-out or cancellation
respectively) at a given 
site as the relevant time scale of the model before crossover to
diffusive Gaussian behavior. 

\section*{Acknowledgments}

One of us (RDW) wishes to thank Sven L\"ubeck for fruitful discussions and 
help on Figure 8 and the MINERVA foundation for financial support.
GMS thanks Deutsche Forschungsgemeinschaft (DFG) for financial support
and the Department of Physics, University of Oxford for kind hospitality.

\
\end{document}